\documentclass[english,pre,showpacs,floatfix]{revtex4}
\usepackage[T1]{fontenc}
\usepackage[latin1]{inputenc}
\usepackage{babel}
\usepackage{epsfig}
\usepackage{amsmath}
\usepackage{amssymb}
\usepackage{epstopdf}
\usepackage{color}

\begin{document}

\title
{\bf Nonlinear response of a linear chain to weak driving}
\author{D. Hennig$^{1}$, C. Mulhern$^{2}$, A.D. Burbanks$^{1}$, and L. Schimansky-Geier$^{3}$
%\footnote{Corresponding author: andrew.burbanks@port.ac.uk Tel:+44(0)2392846458 Fax:+44(0)2392846353.}
}
\medskip
\medskip
\medskip
\affiliation{1 Department of Mathematics, University of Portsmouth,
  Portsmouth, PO1 3HF, UK\\ 2 Max Planck Institute for the Physics of Complex 
  Systems, N\"othnitzer Str.~38, 01187 Dresden, Germany\\3 Institut f\"ur Physik, Humboldt-Universit\"at zu Berlin, Newtonstra{\ss}e
15, 12489 Berlin, Germany
}

\begin{abstract}
\noindent
We study the escape of a chain of coupled units  over the barrier of
a metastable potential. It is demonstrated 
that a very weak external driving field 
with suitably chosen frequency suffices to accomplish speedy escape. The latter requires the  passage 
through a transition state the formation of which is triggered by  permanent feeding of energy from a 
phonon background into humps of localised energy and elastic interaction of the arising breather solutions. In fact, cooperativity between the units
of the chain entailing coordinated energy transfer 
is shown to be crucial for enhancing the rate of escape in an extremely 
effective and low-energy cost way where the effect of entropic localisation and breather coalescence conspire. 
\end{abstract}
%\pacs{05.45.Ac, 05.60.-k, 05.45.Pq, 05.60.Cd}

\maketitle

%\section{Introduction}

Recently there has been increasing interest in the escape of coupled degrees of freedom or chains of interacting  
units out of metastable states \cite{Sung}-\cite{EPL}. Escape is accomplished when the considered object overcomes a potential barrier
separating the local minimum of the potential landscape from a neighbouring domain of attraction. The activation energy 
required to surmount the energetic bottleneck can be provided in different ways. There is the possibility of 
stochastic escape occurring in the presence of a heat bath that is sampled for the optimal fluctuations triggering an 
event of escape. Alternatively, in the noise free situation the energy can be supplied in a single shot under 
micro-canonical circumstances. Spontaneous localisation of energy due to modulational 
instability promotes the formation of 
localised humps on the chain. Escape is connected with the crossing of a saddle point in configuration space. 
In particular, a
sufficient amount of energy needs to be concentrated in an associated critical nucleus called the transition state. 
Adopting the latter and by passing through it the chain is able to surmount the barrier.  In addition the activation 
energy may be 
injected into the system by an external time-dependent field \cite{Maniadis}-\cite{EPL2} and waves \cite{Ruffo}. 
The objective of this manuscript is to elaborate on an escape scenario of a chain of interacting 
units over the barrier of a 
metastable potential promoted by a very weak external periodic field.  
The initially almost homogeneous chain, possessing virtually no energy,  is brought into
the nonlinear regime where it may exhibit  spontaneous energy
localisation. This process is driven by an instability of the homogeneous chain 
with respect to spatial fluctuations triggered by the external field.
Subsequently a localised pattern may form on the chain. 
With this work we  intend to demonstrate that already a  very weak forcing with a 
suitably chosen frequency suffices so
that the necessary energy gets localised in order  that the chain locally adopts and overcomes 
the transition state promoting escape over the barrier.

%\section{The driven oscillator chain}\label{section:model}

We study a one-dimensional lattice of
nonlinear and driven coupled oscillators.
Throughout the following we  use dimensionless parameters, as obtained after appropriate scaling of
the corresponding physical quantities. The coordinate $q$
of each individual nonlinear oscillator with a unit mass
evolves in a cubic, single well on-site potential of the form
\begin{equation}
 U(q)=\frac{\omega_0^2}{2}q^2-\frac{a}{3}q^3;\,\,\,a \geq 0.
\end{equation}
This potential possesses a metastable equilibrium at
$q_{min} = 0$, corresponding to the rest energy $E_{min} = 0$
and exhibits a maximum that is located at $q_{max} = \omega_0^2/a$
with energy $E_{max} \equiv \Delta E = \omega_0^6 /(6a^2)$. Thus, in order for
particles to escape from the potential well of depth
$\Delta E$ over the energy barrier and subsequently into the
range $q > q_{max}$, a sufficient amount of energy needs to
be supplied. The lattice dynamics is governed by the
following system of coupled equations
\begin{eqnarray}
 \ddot{q}_n&+&\omega_0^2q_n-aq_n^2+\kappa [q_{n+1}+q_{n-1}-2q_n]-F(t)=0\,.\label{eq:system}
\end{eqnarray}
The coordinates $q_n(t)$ quantify the displacement of the
oscillator in the local on-site potential $U$ at lattice site $n \in [1, N ]$. 
The oscillators, referred to as units, are coupled
linearly to their neighbours with interaction strength $\kappa$.
A homogeneous external modulation field $F(t)$ globally acts upon the system. 
The field is provided by a periodic monochromatic driving force of amplitude $A$,
frequency $\omega$ and phase $\theta_0$ given as 
\begin{equation}
 F(t)=A \sin(\omega t+\theta_0).\label{eq:periodic}
\end{equation}

We use periodic boundary conditions according to $q_{N +1} = q_1$ and fix the parameters as 
follows: $\omega_0^2 = 2$, $\theta_0=0$, and $a = 1$, yielding $\Delta E = 4/3$. In what follows  we apply a
very weak external periodic field of small amplitude $A=0.003$
which contributes with a value of $\min_{t} F(t) q_{max}= -A q_{max}=-0.006$ to a diminutive lowering of 
the potential barrier only.
 A deterministic 
escape scenario in the conservative, undriven limit of
system (\ref{eq:system}) has been explored in \cite{EPL}.
The system (\ref{eq:system}) has been integrated numerically using a fourth-order Runge-Kutta scheme.
In our simulations the chain consists of $N=100$ units. 
The units are initialised  such that 
the position of all units are randomised around the bottom of the potential $q_0=0$ in the range $|q_n(0)-q_0|<0.1$ 
so that the energy of a unit does not exceed the value 
$E_0=0.01= 0.0075 \times \Delta E$ and hence is negligibly small compared to the barrier energy $\Delta E$. 
%%% coupling energy
The whole chain is thus initialised close to an almost homogeneous 
state, but yet sufficiently displaced in order
to generate nonvanishing interactions, enabling the exchange
of energy among the coupled units.
Starting from a random pattern with units containing vanishingly little energy compared to the barrier energy,
the question then is, 
what is the impact of the very weak periodic driving force regarding escape of the chain over the barrier? 
As the choice of the frequency of the external periodic field 
$\omega$ is concerned  a value  close to the frequency of harmonic oscillations
of the units close to the bottom of the 
potential well, $\omega_0=\sqrt{2}$ seems appropriate since for a start 
phonons need to be excited on the chain. 
The band of phonon frequencies $\omega$ is determined by 
\begin{equation}
 \omega_{ph}^2=\omega_0^2+4 \kappa \sin^2\left(\frac{k}{2}\right),\label{eq:phononband}
\end{equation}
with wave numbers $k \in [0,\pi]$. The frequency of breathers emerging on 
the chain have their frequencies below the lower edge of the phonon band.

For a driving frequency $\omega=1.4 \simeq \sqrt{2}$  the chain rapidly attains
a non-equilibrium steady state of almost 
uniform amplitude, $q_n(t)=q_0(t)$, to which it remains entrained for some time.
This state is in very good agreement with the solution obtained from the linear system,  
which for  zero initial conditions $q_n(0)=\dot{q}_n(0)=0$ is given by
\begin{equation}
 q_n(t)=q_0(t)=\frac{A \omega}{\omega_0^2-\omega^2}
 \left[\frac{1}{\omega}\sin(\omega t)-\frac{1}{\omega_0}\sin(\omega_0 t)\right]\,.\label{eq:driven}
\end{equation}
 Using addition theorems we can express the solution $q_0(t)$ for $\omega_0 \approxeq \omega$ as 
\begin{equation}
 q_0(t)\simeq \frac{2A}{\omega_0^2-\omega^2}\cos\left(\frac{\omega+\omega_0}{2}t\right)
  \sin\left(\frac{\omega-\omega_0}{2}t\right).
 \end{equation}
The (slowly varying) envelope of  $q_0(t)$ attains its maximal amplitude at $t=\pi/(\omega-\omega_0)$. 
At this moment, despite the fact that  the amplitude of the chain has grown already  
(to the brink of the weakly anharmonic regime), the amount of gained energy 
remains too small 
to raise the chain near the potential barrier. 
Nevertheless, as the chain is initialised in a nonhomogeneous state (albeit of low amplitude) 
the units perform small-scale fluctuations around the non-equilibrium 
uniform steady state giving the possibility to render the latter unstable. For near-resonance drivings,
$\omega \simeq \omega_0$, we adopt the 
parameter values such that $v_0=2A/(\omega_0^2-\omega^2) \lesssim 1$ bounding
accordingly the  amplitude of $q_0(t) \simeq v_0 \cos[0.5(\omega+\omega_0)t]\sin[0.5(\omega-\omega_0) t]$. 
While the chain retains the nonequilibrium 
steady state solution $q_0(t)$  with some more growth 
of its amplitude  the weakly nonlinear regime 
is entered (see also below). In the following we perform a stability analysis
of the uniform non-equilibrium steady state solution $q_0(t)$.
To this end we write with respect to spatial perturbations $u_n(t)$:
\begin{equation}
 q_n(t)=q_0(t)+u_n(t).
\end{equation}
For the imposed periodic boundary conditions we express the perturbations as a Fourier-series expansion
\begin{equation}
 u_n(t)=\sum_k \xi_k(t) \exp(ikn)+c.c.,
\end{equation}
with wave numbers $k\in [0,\pi]$ yielding a Mathieu-type equation for the mode amplitudes $\xi_k$
\begin{equation}
 \ddot{\xi}_k+\left\{\omega_{ph}^2 - 2a u_0\cos\left(\frac{\omega+\omega_0}{2}t\right)
  \sin\left(\frac{\omega-\omega_0}{2}t\right)\right\}\xi_k=0,\label{eq:Mathieu}
\end{equation}
where $\omega_{ph}$ is given in Eq.\,(\ref{eq:phononband}) and $u_{0}$ is the amplitude of the nonequilibrium 
steady state solution in the weakly nonlinear regime.

\begin{figure}
\includegraphics[scale=0.25]{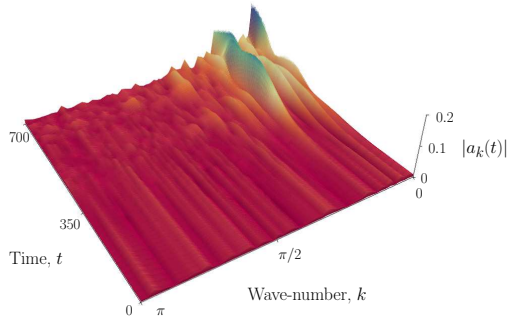}
\caption{Time evolution of the spatial Fourier transform of $q_n(t)$.
The values of the parameters are $\omega_0^2=2$, $a=1$,  $A=0.003$,  
$\omega=1.4$, $\theta_0=0$, and $\kappa = 0.1$.} \label{fig:spatial}
\end{figure}

We investigated the instability region of this Mathieu-type equation in the $A-k$ parameter plane for 
driving frequency $\omega=1.4$.
For $A=0.003$ the position of the bottom
of the instability band is at $k_c\simeq 0.16$  determining the critical unstable
wave number. Thus
the creation of a pattern consisting of localised humps (breathers), with wavelength determined by $\lambda_c = 2\pi/k_c$, 
can be expected. 
In fact, the time evolution of the spatial Fourier-transform, 
\begin{equation}
 q_n(t)=\sum_k a_k(t) \exp(ikn)+c.c.,
\end{equation}
displayed (for one realisation of frozen noise $|q_n(0)-q_0|<0.1$ and $p_n(0)=0$) in Fig.~\ref{fig:spatial},
corroborates this feature. For times $t\gtrsim 400$ (corresponding to $90$ periods 
$T_0=2\pi/\omega_0$ of harmonic oscillations near the bottom 
of the potential well) the chain departs from the regime of nearly uniform steady state, which is
indicated in the spatial Fourier spectrum by the formation of pronounced Fourier components 
in the domain of low wave numbers. 

In accordance with our stability analysis one notices for 
the spatio-temporal evolution of the coordinates $q_n(t)$ that 
due to the  instability of the steady state, and  the ensuing energy exchange 
among the units, after some time small fractions of energy can become localised in humps 
(breathers).
We observe that a pattern evolves in the course of time where for some lattice sites 
the amplitudes grow considerably while remaining small 
in the adjacent regions. That is an array of large wave length (small wave number) chaotic breathers is
formed.
Upon moving, these breathers tend to collide inelastically with others.
In fact, various breathers merge to form larger amplitude breathers, proceeding 
preferably such that the larger amplitude breathers grow at the expense
of the smaller ones. As a result, more energy becomes
strongly concentrated within confined regions of the chain. 
If at least one of the breathers can be sufficiently  strongly amplified on a segment 
of the lattice such that the associated maximal amplitude
grows to the proximity of the barrier level, then a crossing of the energy barrier for this segment becomes achievable. Moreover, in
\cite{EPL} it was shown
that such a localized state might adopt the hairpin shape of the critical
localized mode (transition state) and if the involved amplitudes become overcritical escape is
realised. However, for this to happen, at first one unit has to absorb sufficient
energy to completely surmount the barrier. For our choice of parameter values 
the energy of the transition state amounts to 
$1.35 \times\Delta E$ \cite{EPL}. 
We emphasise that it is the comparatively high mobility of the chaotic breathers
leading to collisions with other (standing and/or moving)
breathers that enhances the concentration of energy. 

\begin{figure}
\includegraphics[scale=0.25]{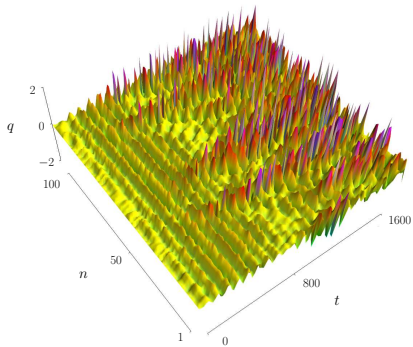}
\caption{Spatio-temporal pattern of the coordinates $q_n(t)$ prior to escape for very weak periodic driving.
The values of the parameters are $\omega_0^2=2$, $a=1$,  $A=0.003$,  
$\omega=1.4$, $\theta_0=0$, and $\kappa = 0.1$. The pattern is shown up to times shortly before the escape of the chain 
takes place.} \label{fig:spatioc}
\end{figure}

In Fig.~\ref{fig:spatioc}
we show a typical spatio-temporal pattern $q_n(t)$ for a very weakly  driven 
chain for the same realisation of frozen noise $|q_n(0)-q_0|<0.1$ and $p_n(0)=0$ as in Fig.~\ref{fig:spatial}. 
A few moving localised humps (chaotic breathers) are discernible. After $\sim 1690$ time units  
the chain adapts locally a hairpin-like configuration 
and by passing through the latter 
with sufficiently large (positive) velocities a subsequent escape of the involved 
units is initiated. Return of the escaped units 
over the barrier into the original potential well is virtually excluded and consecutively all units, 
being pulled by the already escaped ones, 
manage to climb over the barrier. 
Eventually the chain propagates freely beyond the potential barrier with increasing kinetic energy. 
We emphasise that the global escape of the chain relies on appropriate cooperations of the chain units 
leading to enhanced energy concentration -- an effect that is absent for uncoupled units. In fact, 
for vanishing 
coupling $\kappa=0$ none of the periodically driven units escaped during the long simulation time 
$T_s=10^5 \sim 2252 \times T_0=2\pi/\omega_0$.

To quantify the change of the energy of the chain in response to the external driving and internal processes of energy 
redistribution we 
attribute to each unit a site energy
\begin{equation}
 E_n=E_{kin,n}+E_{pot,n}+E_{driv,n}
\end{equation}
where the kinetic site energy, potential site energy, and the driving energy are given by 
\begin{eqnarray}
 E_{kin,n}&=&\frac{1}{2}\dot{q}_n^2\\
 E_{pot,n}&=&U(q_{n}) +\frac{\kappa}{4}(q_{n+1}-q_n)^2+
 \frac{\kappa}{4}(q_{n}-q_{n-1})^2\\
 E_{driv,n}&=&F(t) q_n 
\end{eqnarray}
respectively, and 
compute the average kinetic energy and the average potential energy  of the chain defined as
\begin{eqnarray}
 \bar{E}_{kin}&=&\frac{1}{N}\sum_{n=1}^N \frac{1}{T}\int_{0}^T E_{kin,n}(t) dt\\
 \bar{E}_{pot}&=&\frac{1}{N}\sum_{n=1}^N \frac{1}{T}\int_{0}^T E_{pot,n}(t)dt
% \bar{E}_{driv}&=&\frac{1}{N}\sum_{n=1}^N \frac{1}{T}\int_{0}^T E_{driv,n}(t)dt.
\end{eqnarray}
Due to its diminutive size the contribution of the driving energy $E_{driv}$ can be ignored.  

\begin{figure}
\includegraphics[scale=0.75]{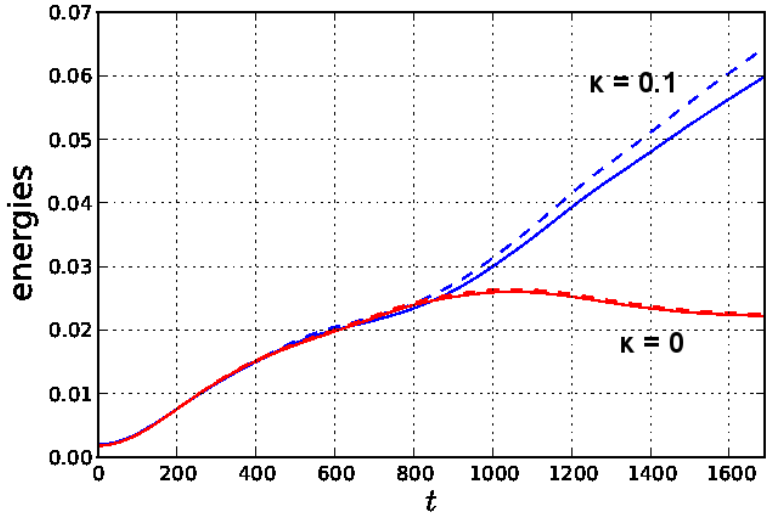}
\caption{Temporal evolution of the average kinetic (solid line)  and average 
potential (dashed line) energy for weak periodic driving for 
coupled units ($\kappa = 0.1$) and uncoupled units ($\kappa = 0$) as indicated in the plot.
The values of the remaining parameters are $\omega_0^2=2$, $a=1$,  $A=0.003$,  
$\omega=1$, and $\theta_0=0$. The time evolution is shown up to times shortly before the escape of the chain 
takes place.} \label{fig:energyp}
\end{figure}

Fig.~\ref{fig:energyp} shows the temporal evolution of $\bar{E}_{kin}$ and  $\bar{E}_{pot}$  contained in the chain of 
coupled units.  
Throughout the time the chain of coupled units gains on average kinetic and potential energy from the external driving field.  
Up to times $t \lesssim 400$,  when the chain is still in the nearly uniform non-equilibrium steady state, 
the average potential and kinetic energy are of equal amount.
 For later times,  when the first humps form on the chain the dynamics enters the weakly nonlinear regime 
 coinciding with the fact that for the ratio of the anharmonic part and harmonic part of the potential exceeds a
 threshold value, viz.  
 \begin{equation}
  \frac{\frac{a}{3}\sum_n |q_n|^3}{\frac{\omega_0^2}{2}\sum_n q_n^2}>0.1.
 \end{equation}
As a hallmark of soft potentials,  the
energy is on average not equally divided between 
potential and kinetic ones, as for a soft oscillator the oscillation frequency decreases and the 
density of states increases with increasing 
amplitude (energy). That is, on average the energy spends relatively more time in potential form 
than in kinetic form.  For comparison we
show also the average energies for uncoupled units, viz. for $\kappa=0$, starting with the same initial 
conditions as the coupled units. Until $t \simeq 400$ the behaviour of coupled units and uncoupled units is 
virtually the same. However for times $t\gtrsim 400$ the first minor deviations occur and later at $t\simeq
850$ 
the average kinetic and potential energy of coupled units increases further  
monotonically while for uncoupled units almost stagnation is found giving evidence that cooperative effects
are crucial for the growth of energy observed for coupled units. 

In this context we also consider the  average temperature of the chain defined as the mean kinetic
energy in a frame co-moving with the 
center of mass of the chain as
\begin{equation}
 T_{chain}=\frac{1}{2} \big\langle (\dot{q}_n(t)-\bar{\dot{q}}(t))^2\big\rangle 
\end{equation}
\begin{figure}
\includegraphics[scale=0.75]{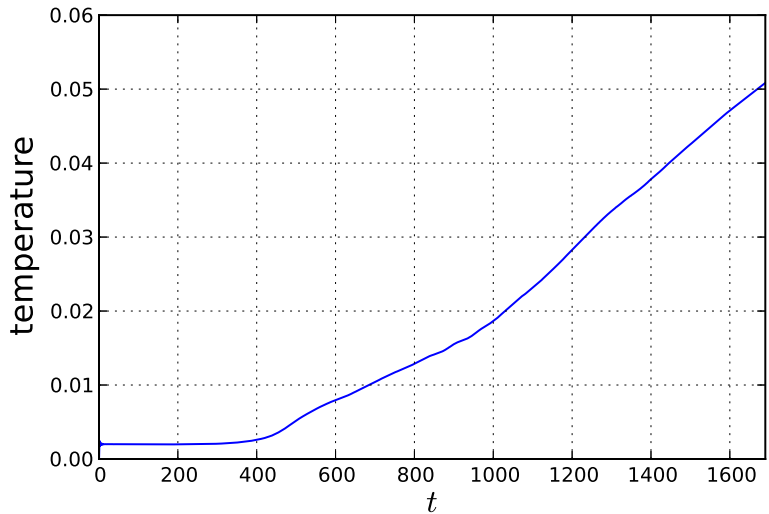}
\caption{Temperature of the chain for weak periodic driving.
The values of the parameters are $\omega_0^2=2$, $a=1$,  $A=0.003$,  
$\omega=1$, $\theta_0=0$, and $\kappa = 0.1$. The time evolution is shown up to times shortly before the escape of the chain 
takes place.} \label{fig:temperature}
\end{figure}

\noindent
where $\bar{\dot{q}}$ is the velocity of the center of mass. Fig.~\ref{fig:temperature} shows the
temporal 
behaviour of the system temperature  up to the moment of 
time when the first unit surmounts the potential barrier initiating global escape. Comparing the 
behaviour of the average
kinetic energy $\bar{E}_{kin}$, and the temperature of the chain $T_{chain}$,  it becomes 
clear that the latter follows 
the former, $T \simeq (0.5-0.8)\times \bar{E}_{kin}$, and thus, comparatively the kinetic energy of the center of 
mass of the chain diminishes as time progresses.

During the time when the chain is still in the  linear regime ($t \lesssim 400$) the temperature remains constant
indicating little variability 
between the units (see above). Upon entering the nonlinear regime related with the emergence  
of breathers the average temperature 
increases monotonically but with a weakly varying rate. In fact, after $1000$ time units (equivalent to $\sim
227$
periods of harmonic 
oscillations near the bottom of the potential well) the slope of the
graph
increases further
indicating a regime of enhanced interactions (`collisions') caused by moving chaotic breathers  traversing the
chain 
(see below). 
With further growth  of the temperature going along with enhanced localisation of energy the chain eventually overcomes  
the potential barrier.
The temperature exhibits a  
stretched exponential  time dependence
\begin{equation}
 T_{chain} \sim \\exp(a t^b)
\end{equation}
with  coefficients $a$ and $b$. An analogous behaviour was found 
in the reverse problem concerning the relaxation dynamics of a lattice chain 
which is initially thermalised and afterwards put in contact with a 
cold bath \cite{GTS96},\cite{Bikaki}.

To gain more insight into the nature of the escape process we consider the 
rate of change of the site-energy
\begin{equation}
 \dot{E}_n=\frac{\kappa}{2}[(q_{n+1}-q_n)(\dot{q}_{n+1}-\dot{q}_n)-(q_n-q_{n-1})(\dot{q}_n-\dot{q}_{n-1})]
+F(t)\dot{q}_n +\dot{F}(t)q_n.\label{eq:change}
\end{equation}
The energy of a unit can be changed via two channels. There is the possibility of energy exchange
with the external field $F(t)$ involving both the velocities $\dot{q}_n$ and coordinates $q_n$ of a unit. 
With proper phase relations 
between the oscillations of a  coordinate $q_n(t)$  
and the derivative of the external periodic field $\dot{F}(t)$ and/or 
between the velocity $\dot{q}_n(t)$ and  the external periodic field $F(t)$
a unit can gain energy during an interval 
$\Delta t$ providing $\int_t^{t+\Delta t} d\tau [\dot{q}_n(\tau)F(\tau)+q_n(\tau) \dot{F}(\tau)]>0$. 
In particular, during the 
initial stage of linear behaviour of the system, no mattter if coupled or uncoupled, its energy 
gain from the external field  is fairly pronounced. This behaviour is characterised by nearly 
in-phase motion of the 
derivative of the external field $\dot{F}(t)$ and the mean value of the coordinates $\bar{q}$ 
whose evolution is determined by Eq.\,(\ref{eq:driven}) as well as in-phase motion of the mean value 
of the velocities $\bar{\dot{q}}(t)$ and the external field $F(t)$. 

Upon entering the anharmonic regime for coupled units a pattern of localised humps arises, which 
naturally is impossible for 
non-coupled units. 
The question is:  What is the mechanism of ongoing energy injection into breathers being responsible
for their observed growth? 
With the onset of spatio-temporal chaos, the almost in-phase coherence between $\bar{q}$ 
and $\dot{F}(t)$ on the one hand, and between $\bar{\dot{q}}$ and $F(t)$ on the other, 
can get temporarily and/or locally lost impeding further energy growth.  Nonetheless there are periods during which the phase correlations 
are maintained so that (at least locally) energy pumping from the external field into the corresponding 
units of the chain is accomplishable.
%Interestingly for $t \gtrsim 625$ the 
% average driving energy $\bar{E}_{driv}$ with non-zero coupling diminishes and at $t \simeq 950$
% falls even below the driving energy with zero coupling (see Fig.~\ref{fig:energyp}). 
% This is the repercussion of  
% entropic localisation inherent to the standing large-amplitude breathers formed on
% the coupled chain biasing effectively
% %constraining considerably 
% the  mean value of the coordinates $\bar{q}(t)=\sum_{n=1}^N q_n(t)/N$ towards the 
% range of positive values (viz. to the range of the soft branch of the potential $U(q)$)  
% whereas $\bar{q}(t)$ of  the uncoupled chain performs permanently symmetric oscillations around 
% zero being more in tune with $F(t)$ than its coupled chain counterpart. 

% As a result the phase relation between the harmonic oscillations of the external field $F(t)$ and
% the anharmonic oscillations of 
% the breathers (oscillating with frequencies below the lower edge of the phonon band)
% adjusts such that temporarily ${E}_{driv}(t)$ 
% becomes negative contributing to the lowering of  $\bar{E}_{driv}$.

%%%does not rule out that the degree of localisation enhances 
% Although the average driving energy  of coupled units 
% reduces steadily there is apparently 

% Nevertheless there is sufficient amount of energy  injected 
%  into parts of the chain  so that subsequent  internal energy redistribution locally increases 
%  the amount of energy stored in breathers 
%  which in turn enhances
%  entropic localisation. 
 
\begin{figure}
\includegraphics[scale=0.75]{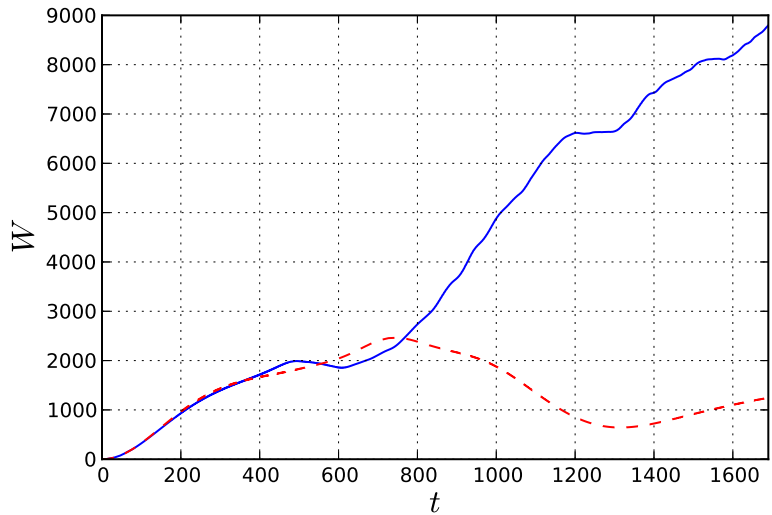}
\caption{Work done on  the chain by the  weak periodic driving for $\kappa=0$ (dashed line) and $\kappa=0.1$ (solid line).
The values of the remaining parameters are $\omega_0^2=2$, $a=1$,  $A=0.003$,  
$\omega=1$, $\theta_0=0$, and $\kappa = 0.1$.} \label{fig:work}
\end{figure}

In addition, the permanent impact of the external periodic field 
serves that itinerant chaotic breathers merge eventually with large-amplitude breathers. In fact, inbetween the few 
large-amplitude standing breathers, the external driving, with 
frequency almost coinciding with the frequency at the lower edge of the phonon band,
creates permanently a phonon `bath' background feeding energy into arising
small-amplitude (chaotic) breathers.   
We emphasise that the cooperativity of the units plays a vital role for achieving escape in such an effective and 
low-energy-cost manner.
To gain more insight into the 
energy gain of the coupled chain we plot in Fig.~\ref{fig:work} the work done by the external field on the chain 
which 
is determined by
\begin{equation}
 W(t)= \frac{1}{N}\sum_{n=1}^N \int \displaylimits_0^{t} F(\tau)\dot{q}_n(\tau)d\tau. 
\end{equation}
During the initial linear regime, characterised by almost in-phase motion of the external field $F(t)$ and the
mean value of 
the velocities $\bar{\dot{q}}$, the work done on the uncoupled units and the coupled ones coincides. However, at 
later times the difference becomes drastic as the work done on the coupled units increases 
in the course of time whereas 
the work done on the uncoupled units is subdued and even diminishes temporarily. For the former this 
means that on average 
the velocities of the units not involved in standing large-amplitude breathers enhance in 
tune with the external periodic field. This instigates motion of small-amplitude chaotic breathers 
(emanating from the phonon background) along the chain
and the energy of the latter
gets absorbed by standing higher-amplitude breathers 
upon coalescence. Furthermore, as long as the standing breathers are of undercritical amplitude, and hence their 
frequency remains close to the driving frequency, 
direct (resonant) injection of energy from the external field into them is possible as a result of which 
the amplitude of the latter grows and its frequency gets shifted further below the phonon band. Eventually for 
overcritical amplitudes the resulting mismatch between the frequency of the breathers and the frequency of the external field 
becomes too great for further direct energy feeding from the external field into the breather. Then the only way for  a
breather to gain more energy is by means of internal energy distribution between the units of the chain.

The internal mechanism of energy exchange along the chain is due to the coupling term 
being responsible for the energy exchange of a unit with its nearest neighbours via the springs connecting the units.
The larger the stress $|q_{n\pm1}-q_n|$ --- the potential energy
stored in the spring --- 
  and/or the velocities $|\dot{q}_{n\pm1}|$, $|\dot{q}_n|$, the higher is the 
rate of internal change of the site-energy. The internal exchange of energy between neighbouring units is blocked
when 
$q_{n\pm 1}=q_n$ or $\dot{q}_{n\pm 1}=-\dot{q}_n$, that is when neighbouring oscillators perform 
equal amplitude in-phase motion 
or out-of-phase motion with respect 
to each other with opposite sign of velocities. 
%Contrarily
Moreover, once a unit has acquired a high energy, it is
retained for a fairly long time due to the fact that in a soft oscillator the energy spends relatively
more time in potential than in kinetic form. The reason is that in soft potentials the oscillation
frequency decreases with increasing amplitude. In conjunction with the fact that the density of
states increases with increasing amplitude, attaining and preserving higher-amplitudes is 
entropically more favourable (see also \cite{Reigada}). 
Thus, during the major part of an oscillation period of
a unit, after it has gained energy from the external field, its neighbours, or impacting moving breathers, 
the displacement of this unit remains
large while the velocity is low. Therefore, 
this entropic localisation mechanism impedes the energy exchange of a higher 
amplitude unit with the surroundings. Conclusively, localisation of energy 
%(resulting in spatial variability) 
minimises the
free energy as it is favoured, with respect to maximisation of entropy, that the energy gaining
units populate regions in phase space where the density of states is higher.

\begin{figure}
\includegraphics[scale=0.75]{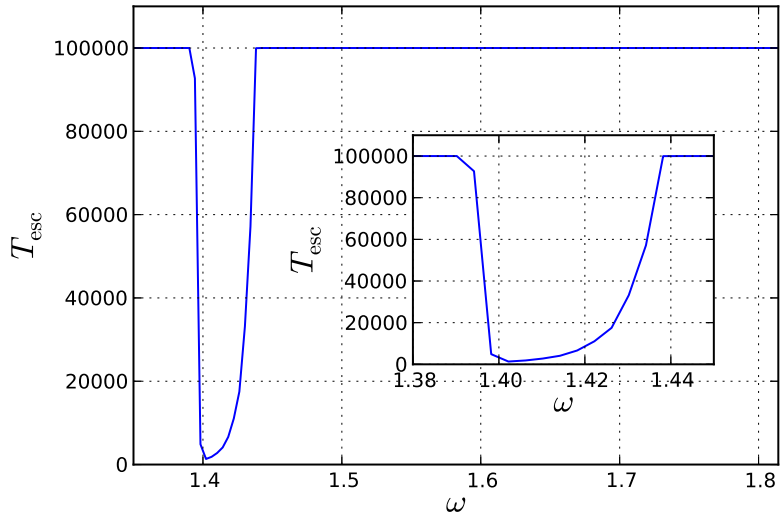}
\caption{Mean escape time of the chain as a function of the driving frequency $\omega$ for a small fixed driving strength $A=0.003$.  
The values of the remaining parameters are $\omega_0^2=2$, $a=1$,  $A=0.003$,  
$\omega=1.4$, $\theta_0=0$, and $\kappa = 0.1$.} \label{fig:escape}
\end{figure}

Finally, we present our results regarding the mean escape time $T_{esc}$ of the chain. 
The escape time for the chain is defined as the
average of the moments at which the $N$ amplitudes of the
escaping units pass the value $q=200= 100 \times q_{max}$  beyond the barrier
location. $T_{esc}$ vs. the driving frequency for a small driving amplitude $A=0.003$ 
is displayed in Fig.~\ref{fig:escape}.
The averages were performed over $100$ realizations of random initial conditions. 
There is a window of frequencies $1.395 \lesssim \omega \lesssim 1.437$ 
for which speedy escape is accomplished and outside of this window  not a single event of escape takes place throughout the simulation time 
$T_s=10^5$. 

In summary, in this study we have investigated the escape problem of a chain of harmonically coupled units over the 
barrier of a metastable potential. Energy is injected into the system by means of an applied external 
time-periodic field.
Notably, even for a very weak driving force we have observed fast escape for a chain 
situated initially extremely close to the 
bottom of the potential well and thus containing a vanishingly small amount of energy.
For a suitably chosen  driving frequency almost coinciding with the frequency at the 
lower edge  of the phonon band of linear oscillations, as a start, an almost uniform oscillating state 
of the chain is excited. The amplitude of the latter rises (slowly) in time and 
upon entering the weakly nonlinear regime the almost uniform state becomes unstable with respect to spatial perturbations. This triggers
the formation of a few localised humps (standing breathers) coexisting with a phonon `bath' background
inbetween them.
Due to the effect of entropic localisation for the standing breathers the process of their energy reduction is impeded. 
Contrarily, the process in the other direction is entropically favoured. That is, due to the fact 
that the driving frequency lies just below the phonon band further resonant energy pumping by the external field into 
standing breathers is possible, provided a proper phase  relation is retained between them. 
In fact, the associated growth of the amplitude of the breathers enhances even entropic localisation.  
However, as with growing amplitude, the frequency of a breather diminishes, and there results a frequency
mismatch 
between the external field and the standing large-amplitude breather hampering  
direct substantial energy feeding from the external field into it. 
Therefore, at this stage the only way a breather can gain more energy is by processes of internal energy redistribution along the chain. 
Conclusively, choosing the frequency of the external driving just below the phonon band is of advantage for two reasons: First, emerging standing breathers can become amplified by 
direct energy gain from the (almost) resonant external field. At the same time, 
externally driving with a frequency almost equal to that of harmonic oscillations near the bottom of the potential well  generates
permanently a phonon `bath' background between the standing breathers forming the source for the emergence of
mobile 
chaotic breathers. Mergence of the itinerant chaotic breathers with standing breathers contributes to their growth. 
Eventually for overcritical amplitudes a standing breather adopts the shape and energy content of the transition state
and by passing through the latter escape of the chain over the barrier gets instigated.

\end{document}